\documentclass[epj]{svjour}
\newif\ifpdf            
\ifx\pdfoutput\undefined \pdffalse \else \pdftrue \fi
\ifpdf 
\usepackage[pdftex]{graphicx}
\pdfcompresslevel9 \DeclareGraphicsExtensions{.jpg,.pdf,.png,.mps}
\else 
\usepackage[dvips]{graphicx}
\DeclareGraphicsExtensions{.eps,.ps} \fi
\usepackage{bm}

\def\vep{\varepsilon}
\newcommand{\beq}{\begin{equation}}
\newcommand{\eeq}{\end{equation}}

\begin{document}

\title{Dynamic Soft Elasticity in Monodomain Nematic Elastomers}
\author{A.~Hotta \and E.M.~Terentjev}
\institute{Cavendish Laboratory, University of Cambridge,
Madingley Road, Cambridge CB3 0HE, U.K. }

\date{\today}

\abstract{We study the linear dynamic mechanical response of
monodomain nematic liquid crystalline elastomers under shear in
the geometry that allows the director rotation. The aspects of
time-temperature superposition are discussed at some length and
Master Curves are obtained between the glassy state and the
nematic transition temperature $T_{ni}$. However, the
time-temperature superposition did not work through the clearing
point $T_{ni}$, due to change from the ``soft-elasticity'' nematic
regime to the ordinary isotropic rubber response. We focus on the
low-frequency region of the Master Curves and establish the
power-law dependence of the modulus $G' \propto \omega^a$. This
law agrees very well with the results of static stress relaxation,
where each relaxation curve obeys the analogous power law $G'
\propto t^{-a}$ in the corresponding region of long times and
temperatures.
} 

\maketitle

\section{Introduction}
For the last two decades, thermotropic liquid crystalline polymers
(LCP) have been a subject of increasing activity due to their
fascinating potential in technical applications, as well as some
fundamentally different properties in comparison to ordinary
liquid crystals. Typical LCP consist of a polymer backbone and
mesogenic groups. The backbone is usually made of siloxane or
acrylic chains and the mesogenic groups are either grafted (end-on
or side-on) using a flexible spacer, thus making a side-chain LCP,
or incorporated directly into the polymer backbone (main-chain
LCP). When the polymer is chemically crosslinked, it forms a
three-dimensional permanent network connected with mesogenic
moieties in what is called Liquid Crystalline Elastomers (LCE)
\cite{fink81}, or gels if the network is swollen by a solvent.
Many new physical results have been reported for LCE, but perhaps
the most important novelty, unique for nematic LCE materials, is
the effect of soft elasticity, where imposed strains could be
completely accommodated without any (or with very low) elastic
response due to the relaxation of internal microstructure. These
unusual mechanical properties of LCE can be directly applied in
technology, using LCE as anomalous damping materials, as well as
artificial muscles or mechanical transducers.

The early research of LCP rheology, focused mostly on the
viscosity of polymers, was summarized in \cite{wissbrun}. Many
interesting results have come to light once the mesogenic groups
were incorporated into linear or branched polymer systems. The
classical Leslie-Ericksen theory of low-molecular-weight liquid
crystals is, in most of the cases, not applicable to LCP systems
because of the underlying polymer dynamics making the response
distinctly non-Newtonian. Hanus et al. \cite{hanus} measured
dynamic-mechanical properties of a weakly crosslinked main-chain
LCP having crystalline, smectic B, smectic A and isotropic phases.
They only observed a clear step in the linear shear modulus $G'$
at the transition temperature from smectic A to the isotropic
state. In the study of LCP having smectic C, smectic A, nematic
and isotropic phases \cite{pakula}, Pakula reported that decreases
in the linear modulus were detected at the smectic/nematic
transition as well as the glass/smectic transition. Also
interestingly, although it was not discussed theoretically at that
time, the \underline{increase} of the modulus could be seen during
the transition from the nematic to the isotropic phase. The issues
of time-temperature superposition in LCP rheology have been
actively discussed by several groups such as Colby et al.
\cite{colby}, Gallani et al.  \cite{gallani2,gallani96},
Fourmaux-Demange et al. \cite{fourmaux-demange} and Weilepp et al.
\cite{weilepp2}. All of these works also reported the
characteristic power-law frequency dependence of the storage
modulus $G'(\omega)$. Colby et al. used a sample possessing three
phases (isotropic, nematic and smectic A) and concluded that the
viscoelastic response is very sensitive to smectic-nematic and
smectic-isotropic transitions, but insensitive to the
nematic-isotropic transition. The time-temperature superposition
worked only at the nematic-isotropic transition. Approaching the
glass transition, $G'(\omega)$ obeys a power law of $G' \propto
\omega ^{0.75}$. The authors suggested that this power law might
be connected to the large polydispersity of their sample
\cite{colby}, which was also argued by Fourmaux-Demange et al.
\cite{fourmaux-demange} for their different LCP system. Gallani et
al. measured dynamic-mechanical response of several LCP materials
and also found a power law behaviour of the storage modulus $G'$,
with the exponent of $0.44$, $0.6$, $0.67$ or $0.75$ depending on
the sample. Here again, the time-temperature superposition worked
only during the nematic-isotropic transitions. This was attributed
\cite{fourmaux-demange,gallani96} to the transient elastic
clusters of macroscopic size (a few tens of micrometers), floating
in a viscous medium. One may recall that a classical Rouse
dynamics of polymer chains would lead to a square-root dependence
$G' \propto \omega^{0.5}$ \cite{Doi}.

In this paper we shall find that the power law dependence is also
effective in crosslinked LCE materials. In their parallel study of
LCE, Gallani et al. and Weilepp et al. \cite{gallani96,weilepp2}
also clearly stated that the rheological response did not change
significantly when crossing an isotropic-nematic phase boundary,
whereas there has been a crucial change at the nematic-smectic A
transition. The power-law frequency dependence of the elastic
moduli has again been reported, with $G'\propto \omega^{0.3}$ in
the smectic A phase. In the isotropic phase, the exponent was
found to be 0.5, similar to the Rouse model prediction. A more
detailed recent study of nematic LCE rheology by Stein et al.
\cite{stein01} also reported $G'\propto \omega^{0.5}$ in a
polydomain nematic elastomers, but could not obtain Master Curves
for monodomain, aligned nematic materials. Working with such
monodomain nematic rubbers, Clarke et al.
\cite{clarkePRL01,clarke_hotta2} reported a significant difference
in response between polydomain samples (where the nematic director
is disordered on the scale above few microns) and monodomain
nematic rubbers (with the director is macroscopically aligned
along the shear direction, or perpendicular to it). A substantial
decrease in the low-frequency storage modulus $G'$ was registered
as the temperature was decreasing through nematic-isotropic
transition. This unusual effect has been theoretically described
by a continuum theory of linear viscoelastic response in oriented
monodomain nematic elastomers \cite{terentjev01}.

Continuous stress relaxation is a related subject, which offers a
different angle of approach to the complex problem of polymer
dynamics exploring the time, rather than the frequency domain.
Chasset and Thirion, in their classical paper \cite{chasset},
described a power law decay in ordinary crosslinked polymer
networks such as the isoprene rubber (IR) and the
butadiene-styrene copolymer rubber (SBR). They found that, after
application of a constant strain, the stress response decreases
with time, reaching at long times a characteristic power-law
regime $\propto 1/t^{0.1-0.15}$ at very long times. In the linear
regime (at small enough deformations), in an incompressible
material (when the extension modulus is simply three times the
shear modulus $G$), this translates directly into the relaxation
of the effective modulus $G(t)$. Many subsequent experiments on
stress relaxation, performed by different groups, have supported
this long-time power law decay with a small exponent. Theoretical
analysis of \cite{curro_pincus}, applying the reptation ideas for
a network with long free dangling chain ends, has shown that the
retraction of these chains could account for the observed stress
relaxation law $\sigma \propto t^{-a}$ with a small exponent $a
\sim 0.1-0.15$.

In liquid crystalline networks, the early results shown a
different response. The problem in such experiments is always the
need to access very long times, when the main transient processes
in the polymer network have relaxed and only the slowest mode
remains. A peculiar dual regime of long-time relaxation in a
polydomain siloxane nematic LCE has been reported by Clarke et al.
\cite{clarkePRL}. A characteristic time $t^{*}{\sim} 3000$\,s
separated a regime of fast power-law relaxation, $G \propto
{t}^{-0.5}$ or higher, and a very slow relaxation at the later
stage, fitted in \cite{clarkePRL} with an inverse-logarithmic law
${\sigma} {\sim} [1+\ln(t/t^*)]^{-1}$. An even faster power-law
decay was reported by Hotta et al. \cite{hotta_terentjev2} in
their study of acrylic nematic LCE, $G \propto t^{-0.67}$ in the
early stage of stress relaxation. At very long times, they find
the different slow power law decay with the exponent $\sim$0.15,
similar to the one observed in natural crosslinked rubber by
Chasset and Thirion.

From this brief summary, it is clear that the power-law dependence
of linear response function on time or frequency is a recurrent
feature of LCE rheology. On the other hand, it is equally clear
that the response is non-universal and the particular features of
the response depend on the material composition and also on the
nematic director texture and orientation with respect to the
applied strain. In this paper we describe a simultaneous study of
dynamic-mechanical and stress-relaxation responses for the two
well-characterised nematic LCE materials. One sample is a
``classical'' siloxane side-chain nematic rubber first prepared
and described in great detail by Finkelmann et al. and others
\cite{kupfer,kupfer2,stein01,clarkePRL01}. The other material had
a high content of main-chain nematic polymer strands, which makes
the effective chain anisotropy very high \cite{Loft}, but also
leads to a very slow dynamics. We shall examine in some detail the
issues of time-temperature superposition, which is necessary to
extrapolate the response function into the regions of time or
frequency not practically accessible on experiment. We find that
the results of two such different experimental techniques are, in
fact, in a good agreement: the observed frequency power laws $G'
\propto \omega^a$ are matched by the corresponding relaxation laws
$G \propto t^{-a}$, if the corresponding ranges of time and
frequency are examined.

\section{Material Characterisation}

\begin{figure}
\centering \resizebox{0.45\textwidth}{!}{\includegraphics{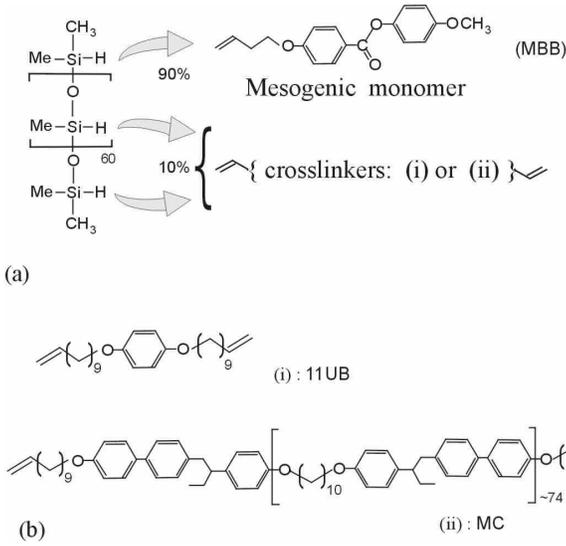}}
 \caption{Schematic illustration of the materials used in this
work. (a) Siloxane backbone chain with Si-H groups reacting with
90~mol\% mesogenic phenyl-benzoate side groups, MBB and 10~mol\%
of di-vinyl crosslinking groups: (b) flexible small-molecule 1,4
alkeneoxybenzene, 11UB (resulting in the SiH material), and the
main-chain nematic polymer of 1-biphenyl-2-phenyl butane, MC
(giving the SiMC material). } \label{chem2}
 \end{figure}

Two nematic elastomers were prepared in our group following the
synthetic procedure pioneered by Finkelmann et al. \cite{kupfer}.
The polymer backbone was a standard poly-methylhydrosiloxane with
approximately 60~$Si$-$H$ units per chain, obtained from ACROS
Chemicals. The pendant mesogenic groups were 4'-methoxy
phenyl-4-(1-buteneoxy) benzoate (MBB), as illustrated in
Fig.~\ref{chem2}, attached to the backbone end-on via the
hydrosilation reaction in the presence of a commercial platinum
catalyst COD, obtained from Wacker Chemie. The first monodomain
nematic elastomer was prepared following the two-step crosslinking
reaction \cite{kupfer}, with 1,4 di(11-undecene) benzene (11UB), a
small flexible difunctional molecule deemed to have relatively
minor effect on the overall mesogenic properties of the liquid
crystalline polymer (synthesized in the house).

The second network was prepared following the similar procedure,
but using di-vinyl terminated polymer chains of
$\alpha$-\{4-[1-(4'-\{11-undecenyloxy\} biphenyl)-2-phenyl]
butyl)-$\omega$-(11- undecenyloxy) poly-[1-(4-
oxydecamethyleneoxy)- biphenyl -2-phenyl] butyl (MC) that
themselves make a main-chain LCP \cite{percec}. In our case, the
crosslinking chains had $\sim$75 rod-like monomers between the
terminal vinyl groups (determined by GPC, polydispersity $\sim$2).
As a result, the properties of a nematic LCE are dominated by
these long main-chain LCP strands connecting the siloxane
side-chain molecules (as the Table~1 indicates, the overall MC
content is $\sim$58\%). In both cases, the crosslinking density
was kept at 10 mol\% of the reacting bonds in the siloxane
backbone, so that on average, each siloxane chain has nine pendant
mesogenic groups between crosslinking sites.

Both elastomers were prepared from the same batch of backbone and
side-group mesogenic materials, with the same relative
concentration of crosslinking groups.

Table~1 summarizes chemical and thermal characteristics of both
materials. Equilibrium transition temperatures were determined on
a Perkin Elmer Pyris 1 differential scanning calorimeter (DSC),
extrapolating to low cooling rates, and the nematic phase
identified by polarizing optical microscopy and X-ray scattering.
The average degree of chain anisotropy was identified by observing
the uniaxial thermal expansion of LCE samples as a function of
temperature. Thermal expansion measurements were made by
suspending the samples, without load, in a glass-fronted oven and
measuring the variation in natural length of the samples with
temperature, $L(T)$; see \cite{Loft} for further details.

\begin{table}
\centering
\begin{tabular}{|l|c|c|c|c|c|c|}
Sample  & 11UB & MC & SC wt & $r$ & $T_g$ &  $T_{ni}$  \\
\hline
SiH & 10\% & 0\% & 87\% & 1.5 & 3${}^{\rm o}$C & 86${}^{\rm o}$C \\
SiMC & 9\% & 1\% & 42\% & 2.8 & 2${}^{\rm o}$C  &  107${}^{\rm o}$C  \\
\hline
\end{tabular}
\caption{Proportions (in mol\%) of crosslinkers 11UB and MC in the
overall crosslinking composition (of the fixed total of 10\%), the
corresponding volume fraction of side-chain mesogenic groups (in
wt\%), the average chain anisotropy $r$ at room temperature,
temperatures of glass and nematic-isotropic transitions. The glass
transition temperatures are approximate, with an error of at least
$\pm 5^{\rm o}C$. \label{tab1} }
\end{table}

\section{Dynamic Mechanical Analysis}

\begin{figure}
\centering
 \resizebox{0.45\textwidth}{!}{\includegraphics{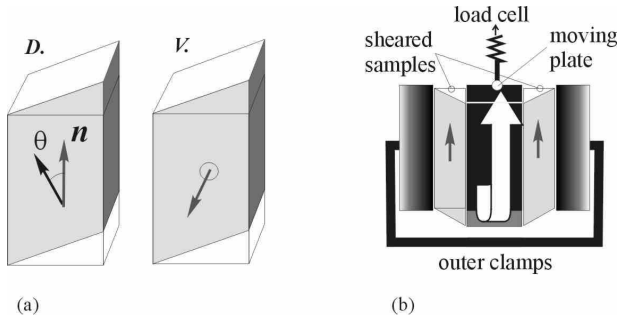}}
 \caption{(a) The geometry of simple shear with two
principal orientations of the initial director $\bm{n}$, labelled
$\bm{D}$ (for $\bm{n}$ along the shear displacement) and $\bm{V}$
(vorticity). No director rotation, hence no soft elasticity,
occurs in the $\bm{V}$-geometry \cite{clarkePRL01}. (b) The
symmetric sample mounting in the DMA device, in the
$\bm{D}$-geometry. See text for the sample dimensions used in
different setups.} \label{geom}
\end{figure}

In this paper, we used the dynamic mechanical apparatus (DMA)
Viscoanaliser VA4000, from Metravib RDS. In the previous work
\cite{clarke_hotta2,clarkePRL01} we have already examined the
anisotropy of the shear response depending on the orientation of
the nematic director, see the sketch in Fig.~\ref{geom}(a). Only
the rotation-inducing geometry reflects the effects of soft
elasticity and here we concentrate on this particular case. The
way in which the samples were mounted in the DMA is shown in
Fig.~\ref{geom}(b). It was important to choose the correct
mounting for a given temperature range because of the great
variation in the magnitude of response between the low-temperature
(or high-frequency) glassy regime and the high-temperature rubber
plateau. We used two different ways of sample mounting: The more
stiff ``sandwich'' geometry, with each sample having the large
area of contact with the clamps and the small thickness, was used
to collect weaker signals from softened samples at higher
temperature (in our DMA setup the thin-film sample had a round
shape). The ``tape'' mounting geometry, in which the rectangularly
cut thin-film sample is clamped sideways with only the small
cross-section area, results in smaller forces and allows to
measure the modulus of very rigid samples up to the glass plateau.
The Table~2 gives the dimensions of typical sample mounting.

A geometry of uniaxial extension, more commonly used in studies of
equilibrium stress-strain relaxation in rubbers, is less
appropriate for an oscillating regime because of slow relaxation
and incomplete sample recovery on each cycle. In the chosen simple
shear setup, the middle part of the sample holder is forced to
move in an oscillating fashion, maintaining the small shear strain
-- between $0.003$ and $0.004$ (that is, 0.3-0.4\%). This value
was chosen from a separate strain-ramping experiment, finding the
range of strains that still return a linear response in the whole
of the chosen frequency range.

\begin{table}
\begin{tabular}{|l|c|c|c|c|}
        &   \multicolumn{2}{c|}{Sandwich}   &   \multicolumn{2}{c|}{Tape}   \\
\hline
 & Thickness  & Area  & Thickness  & Area    \\
\hline
SiH & 0.25~mm &  79~mm${}^2$ & 2~mm & 1.9~mm${}^2$ \\
    &      & $(\pi \times 5^2)$ &   &  $(7.4\times 0.25)$ \\
SiMC & 0.15~mm &  36~mm${}^2$ & 2~mm & 1~mm${}^2$ \\
    &      & $(\pi \times 3.4^2)$ &   &  $(6.7\times 0.15)$\\
\hline
\end{tabular}
\caption{Samples and their dimensions for the simple shear
experiment in two mounting geometries, see text. \label{tab2} }
\end{table}

\subsection{Temperature Scans}

\begin{figure}  
\centering
\resizebox{0.47\textwidth}{!}{\includegraphics{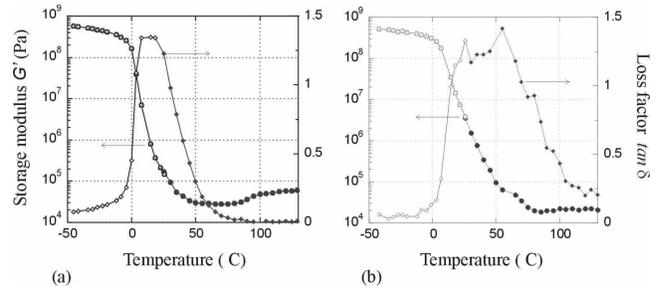}}
\caption[]{Storage modulus $G'$ (circles) and loss factor
$\tan~\delta$ (diamonds) as functions of temperature at
frequencies of 0.01 (a) and 100~Hz (b). The graphs combine the
data obtained from the tape (open symbols) and the sandwich
(filled symbols) sample mounting geometry. Results for SiH. }
\label{SiHtemp}
\end{figure}

\begin{figure}  
\centering
\resizebox{0.47\textwidth}{!}{\includegraphics{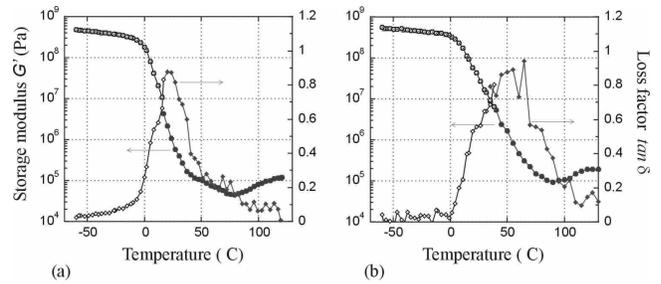}}
\caption[]{Storage modulus $G'$ (circles) and loss factor
$\tan~\delta$ (diamonds) as functions of temperature at
frequencies of 0.01 (a) and 100~Hz (b). The graphs combine the
data obtained from the tape (open symbols) and the sandwich
(filled symbols) sample mounting geometry. Results for SiMC. }
\label{SiMCtemp}
\end{figure}

The materials were studied over a wide temperature range
encompassing the glass and the nematic/isotropic transitions.
Typically the temperature scans ranged from $-50^{\rm o}C$ to
$130^{\rm o}C$, at a number of fixed frequency values.
Figure~\ref{SiHtemp} shows the results for the storage modulus
$G'$ at just two frequencies, 0.01 and 100~Hz, for the SiH sample.
The storage modulus $G'$, which is the real part of the linear
complex modulus $G^*(\omega)$ \cite{ferry} and the loss factor
$\tan~\delta = G''/G'$ (with $G''$ the imaginary part of
$G^*(\omega)$ --  the loss modulus)  were plotted against the
temperature. As we mentioned above, it is technically difficult to
obtain a clean mechanical signal from the sample that undergoes a
change in stiffness of up to $10^4 \rightarrow 10^9$Pa. Therefore,
we used the two ways of mounting the samples, the ``sandwich'' at
high and the ``tape'' at low temperatures. One can see in
Fig.~\ref{SiHtemp}, as well as in Fig.~\ref{SiMCtemp} for the SiMC
material, that the two portions of the full curve, for the tape
and sandwich modes, are connecting quite successfully without any
additional adjustment (the data were discarded if the vertical
error was reaching $15\%$, on the assumption of a mounting fault).
This indicates good reliability of DMA measurements for both the
storage modulus $G'$ and loss factor $\tan~\delta$.

The glass plateau modulus for both SiH and SiMC materials reaches
up to nearly $10^9$~Pa, which is a reasonable value in polymeric
materials. Between $0^{\rm o}C$ and $10^{\rm o}C$, both SiH and
SiMC undergo the glass/nematic transition, showing a dramatic
decrease in $G'$ on cooling, and the associated steep rise in
$\tan~\delta$. The storage modulus decreases down to $\sim
10^5$~Pa which is an expected order of magnitude in rubbery
phases. However, both materials show a substantial drop in $G'$ in
the nematic phase, below $T_{ni}$. This effect, most pronounced at
low frequencies, is the dynamic soft elasticity: the reduction in
the modulus and the rise in internal mechanical dissipation caused
by the underlying director rotations induced by shear
\cite{terentjev01,clarkePRL01}. The relative depth of the drop in
$G'$ is higher in SiMC, the material with a higher chain
anisotropy due to its main-chain LCP component. Above $T_{ni}$,
the storage modulus returns to its (higher) value characteristic
of ordinary isotropic rubbers. At very low frequencies $G'$ is
slightly rising with temperature, revealing that the sample is in
a crosslinked rubbery state with its equilibrium shear modulus $G
\sim n_x k_BT$ (where $n_x$ is the crosslink density and $k_B$ is
Boltzmann factor).

\subsection{Frequency Scans}

\begin{figure}  
\centering
\resizebox{0.47\textwidth}{!}{\includegraphics{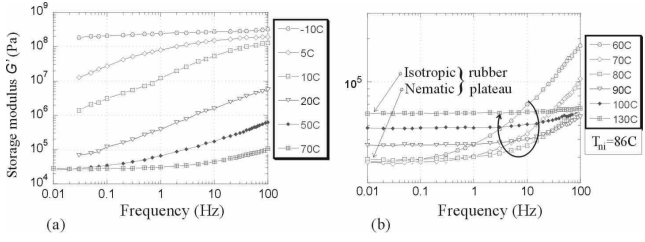}}
\caption[]{Storage modulus $G'$ as a function of frequency at
temperatures ranging from $-10^{\rm o}C$ to $130^{\rm o}C$,
labelled on plots; results for SiH. Plot (a) shows the set of
lower temperatures, including the glass and the nematic rubber
plateau regimes and clearly suggesting the time-temperature
superposition. Plot (b) shows the higher temperatures, below and
above $T_{ni}$, indicating the rise in the rubber plateau value in
the isotropic phase: the dynamic signature of soft elasticity. }
\label{SiHfreq}
\end{figure}

\begin{figure}  
\centering
\resizebox{0.47\textwidth}{!}{\includegraphics{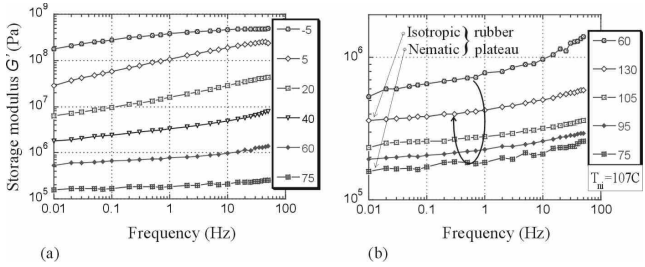}}
\caption[]{Storage modulus $G'$ as a function of frequency at
temperatures ranging from $-5^{\rm o}C$ to $130^{\rm o}C$,
labelled on plots; results for SiMC. Plot (a) shows the set of
lower temperatures, including the glass and the nematic rubber
plateau regimes and again suggesting the time-temperature
superposition. Plot (b) shows the higher temperatures, below and
above $T_{ni}$, same as in Fig.~\ref{SiHfreq}(b). }
\label{SiMCfreq}
\end{figure}

Figures~\ref{SiHfreq} and \ref{SiMCfreq} show the same data for
the storage modulus $G'$, but this time plotted against frequency
at a set of constant values of temperature. In our experiments,
the frequency was changing between 0.01 to 900~Hz, measured 45
points in total. However, we were obliged to discard higher
frequencies: the DMA data sets must pass a series of tests on
internal consistency and failing these tests indicates one of
several possible problems with acquisition (most common is the
internal resonance of the mechanical frame). Only the reliable
range of results is presented in Figs.~\ref{SiHfreq} and
\ref{SiMCfreq}.

The range of storage modulus variation with frequency matches well
with the temperature scans, from $10^9$~Pa down to $10^4-10^5$~Pa.
As in the temperature scans, the average value of $G'$ is rising
on heating, when each sample approaches its isotropic rubbery
phase and the mesogenic effects weaken. In each of the figures,
the plot (b) shows the expanded version of the high-temperature
results, clearly indicating the effect of dynamic soft elasticity.
In addition, one can see for the SiMC material that the
low-frequency rubber plateau is not achieved even at highest
temperatures: the modulus continues to decrease with decreasing
frequency indicating the on-going mechanical relaxation processes
in the sheared rubbery network. It is in contrast with the SiH
results, where one can clearly identify the frequency-independent
rubber plateau in both the nematic and the isotropic phases.

Below we shall discuss this extremely slow relaxation of the
MC-containing elastomers and match these results with the findings
of static stress relaxation. We attribute this slowing down to the
hairpin folds of the backbone of rods with flexible spacers, which
has been demonstrated \cite{elias} to freeze the dynamics even of
non-crosslinked polymer melts.

\subsection{Time-Temperature Superposition}

The frequency scans presented in Figures~\ref{SiHfreq}(a) and
\ref{SiMCfreq}(a) invite the time-temperature superposition
\cite{ferry}. At the same time, one immediately discovers from the
corresponding (b)-plots that such superposition will not be
possible as the monodomain nematic rubber sheared in the ``soft''
$\bm{D}$-geometry undergoes the nematic transition. We assert that
this conclusion can only be made for materials which show a clean
and pronounced soft decrease in $G'$: in a frustrated polydomain
elastomer, or in an overcrosslinked network, the required nematic
director rotation may be impeded by internal constraints and,
therefore, the characteristic nematic soft elasticity masked. In
that case, one might expect the ordinary time-temperature
superposition to be valid, same as in ordinary isotropic polymer
systems.

\begin{figure}  
\centering
\resizebox{0.47\textwidth}{!}{\includegraphics{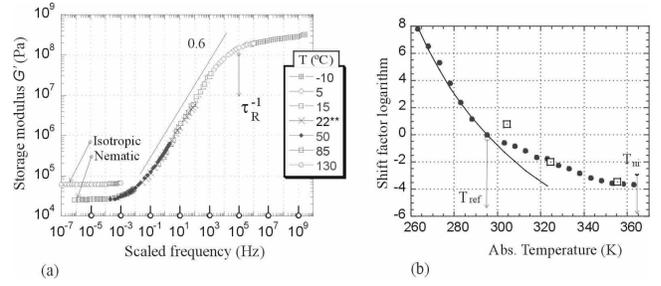}}
\caption[]{Time-temperature superposition of frequency scans for
SiH, at different temperatures.  (a) The Master Curve built at
$T_{\rm ref}=22^{\rm o}C$, superposing the data at temperatures
labelled on plot. Annotations also show the expected Rouse time
and the power-law exponent of the approach to the glass
transition. \ (b) Logarithm of shift factors plotted against
temperature. The curve was fitted with the WLF eq.~(\ref{aT}) at
lower temperatures. Square symbols refer to the discussion of
Section~4 below.} \label{time-temperature1}
\end{figure}

\begin{figure}  
\centering
\resizebox{0.47\textwidth}{!}{\includegraphics{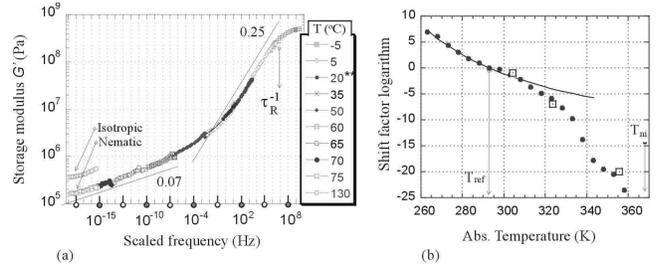}}
\caption[]{Time-temperature superposition of frequency scans for
SiMC, at different temperatures.  (a) The Master Curve built at
$T_{\rm ref}=20^{\rm o}C$, superposing the data at temperatures
labelled on plot. Annotations also show the expected Rouse time
and the power-law exponents. The plateau level is not reached even
at such low extrapolated frequencies. \ (b) Logarithm of shift
factors plotted against temperature. The curve was fitted with the
WLF equation at lower temperatures. Square symbols refer to the
discussion of Section~4 below.} \label{time-temperature2}
\end{figure}

Time-temperature (or, in this case, frequency-temperature)
superposition is a very interesting, practically useful, but also
theoretically ill-defined approach to data analysis in dynamic
mechanical measurements. The empirical algorithm proposed by
Williams, Landel and Ferry (WLF) \cite{ferry} allows extrapolating
the results into the time or frequency ranges far beyond
experimentally accessible. As traditional in this field, we
attempted such a superposition of our data for nematic elastomers.
The storage moduli $G'$ taken in the series of frequency scans at
different temperatures are shifted along the frequency axis until
each data sets superimposes with the previous (with no additional
vertical shifting).  The multiplicative shift factor, $\omega' =
a_T \omega$, is strongly temperature dependent. Another important
factor is the choice of reference temperature $T_{\rm ref}$ at
which the real frequency values are taken. WLF suggested using the
glass transition temperature as $T_{\rm ref}$, however, we decided
against this for the following reasons. First of all, the
presented results show that the dynamic glass transition is, in
fact, difficult to pinpoint as the transformation occurs over a
range of temperatures. Secondly, no additional information is
extracted from the choice of $T_{\rm ref}$ anyway; all it affects
is the overall shift of the Master Curve along the frequency axis.
Our choice of $T_{\rm ref}$ as the ambient room temperature was
dictated by the possibility of determining the key parameter of
polymer chains, the Rouse frequency $\omega_{\rm R}=\tau_{\rm
R}^{-1}$ at the relevant $T_{\rm ref}$, from the position of the
dynamic glass transition on the frequency axis. In both materials,
we find the Rouse time $\tau_{\rm R}\sim 10^{-5}-10^{-6}$s (which
of course must be regarded as a very crude estimate).

Figures~\ref{time-temperature1}(a) and \ref{time-temperature2}(a)
show the results of superposition for the SiH and SiMC samples,
respectively, in the form of the Master Curves. The procedure is
only applicable at the temperatures below the nematic transition
$T_{ni}$. Each graph also shows one data set $G'(\omega)$,
appropriately shifted, at a high temperature, in the isotropic
phase. The values are much higher than on the low-frequency
nematic plateau and, of course, no shifting along the
$\omega$-axis would superimpose them.

The very existence of Master Curves indicates that there is a
unique physical process of mechanical relaxation, which spans the
range of times, or frequencies that is far beyond experimentally
accessible. Our measurements, at each temperature, are only able
to detect a small portion of the whole process, but the
time-temperature superposition suggests a way of building the
whole picture from small pieces. An important part of this
analysis is the temperature dependence of shift factors $a_T$,
which is plotted in Figs.~\ref{time-temperature1}(b) and
\ref{time-temperature2}(b). The empirical WLF relationship is
based on the ideas of glass transition and has been successful in
describing the properties of simple homopolymer melts:
\begin{equation}
\log ~ a_T = -\frac{C_1 (T-T_{\rm ref})}{C_2 + (T-T_{\rm ref})},
\label{aT}
\end{equation}
where $C_1$ and $C_2$ are the constants, called the WLF
coefficients. (In more complex polymers, and in crosslinked
networks, deviations from this equation have been substantial). In
our case, Figures~\ref{time-temperature1}(b) and
\ref{time-temperature2}(b), the shift factors follow the WLF
relation very well at lower temperatures (below $T_{\rm ref}$),
with the  WLF coefficients $C_1 \approx 17.4$ and $C_2 \approx
102$ for both materials,\footnote{Recall that both SiH and SiMC
have been analysed with nearly the same reference temperature,
around $20^{\rm o}$C.} very close to the values found in ordinary
isotropic polymer systems. This is perhaps not surprising when one
recalls that the empirical WLF relationship is ``designed'' to
describe the dynamic glass transition region.

However, the shift factors deviate from the classical WLF behavior
very strongly as soon as the higher temperatures (and,
correspondingly, long relaxation times) are considered. Here, the
two materials behave quite differently: SiH only allows the
superposition of frequency scans with much smaller shifts $a_T$,
while the Master Curve for SiMC demands much greater shifts. One
should note that, in practice, the procedure of shifting of plots
along the frequency axis to achieve superposition is very
sensitive to the value of each shift factor $a_T$ and,
accordingly, the errors in Figs.~\ref{time-temperature1}(b) and
\ref{time-temperature2}(b) are very small. It is more likely that
the noise in the $a_T$ data is due to uncertainties in DMA
temperature or modulus measurements.

Finally, we note that the region of the dynamic glass transition,
in both Master Curves for SiH and SiMC, follows a clear power law
$G' \propto \omega^a$. The plots indicate the non-universal values
of the exponent, similar to many previous observations. The very
low frequency (long-time) tail of continuing relaxation in SiMC
may appear to follow another power law with a much smaller
exponent of $0.07$ also labelled in
Fig.~\ref{time-temperature2}(a). However, we are inclined to
believe that this is an ambiguity of the complex transient regime:
in the previous studies of static stress relaxation
\cite{clarkePRL,hotta_terentjev2} we were distinctly unable to
distinguish between a very low power-law exponent and an even
slower logarithmic relaxation regime. Since the material never
reached the equilibrium, whether in a real experiment or in
extrapolation, the identification of the longest relaxation regime
remained impossible.

\section{Static Stress Relaxation}

Stress relaxation curves were measured at several different
temperatures at a constant extensional strain of $0.09$ for both
SiH and SiMC, following the simple technique described in
\cite{clarkePRL,hotta_terentjev2}. This strain is higher than the
shear strain used in DMA experiments (for technical reasons we
simply could not impose a smaller extension) and the question of
whether the response is still linear may be raised. We have to
stress two points. First of all, the strain of $\sim$9\% is, in
fact, on the border of the initial linear regime of stress-strain
curves measured for both SiH and SiMC, cf. \cite{clarke_hotta1}.
This, however, has to be regarded with caution because it is
impossible to obtain a truly equilibrium stress-strain relation (a
certain rate of strain is always involved) and any extrapolation
to zero strain rate would involve a theoretical model, which may
or may not be ultimately successful. However, as a second point,
we have observed that the key conclusion of this section, about
the power laws of stress relaxation, remains valid even for higher
strains -- on and above the soft plateau when we are certain that
the stress-strain is non-linear. Therefore, in practice, the
choice of the strain step for the static relaxation experiments
was not very important.

\begin{figure}  
\centering
\resizebox{0.47\textwidth}{!}{\includegraphics{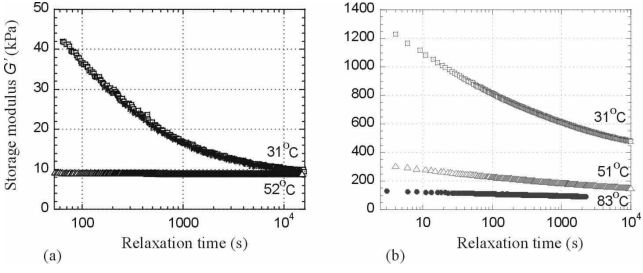}}
\caption[]{Typical stress relaxation curves for the SiH (a) and
SiMC (b) samples, after a step-strain of 9\% is applied
perpendicular to the nematic director (log-time axis). In each
plot, the data sets for different temperatures are labelled. }
\label{stress relaxation curve}
\end{figure}

For the constant-strain  extension setup, each sample was cut with
a razor blade into a rectangular shape of approximately $6\times
2.5\times 0.2\, \hbox{mm}$, measured by a micrometer (narrow
elongated strips minimize the edge effects during the extensional
deformation). The initial sample cross-section area of
$0.5\,\hbox{mm}^2$ was used to calculate the nominal stress
$\sigma$, in units of (mN/mm${}^2 \, \equiv$~kPa), from the force
measured by the temperature-compensated dynamometer (UF1/AD20 from
Pioden Controls Ltd). The values of force, obtained in arbitrary
units, were calibrated with weights at different experimental
temperatures of 30, 50, 80, 105$^{\rm o}C$. The samples were then
extended with the engineering strain $\vep =(L-L_0)/L_0$,
increasing from 0 to 0.09 within a second. Both monodomain
samples, SiH and SiMC, were stretched perpendicular to their
uniform director orientation so that a soft-deformation geometry,
comparable to the $\bm{D}$-shear geometry of DMA experiments, is
achieved. In each case the stress relaxation was measured for two
days $(\sim 10^{5}\hbox{s})$. Figure~\ref{stress relaxation curve}
shows typical stress relaxation curves taken in the series of
these measurements. The data were then fitted with several
possible stress relaxation equations, such as the power-law, the
inverse logarithmic law and the exponentials.

In Fig.~\ref{stress relaxation curve}(a), for the ``faster'' SiH
material, one can note that the stress relaxation was substantial
at a lower temperature of 31${}^{\rm o}C$ and is almost
unnoticeable at higher temperatures (in fact, the data set for
82${}^{\rm o}C$ is not shown because it is hard to distinguish
from the 52${}^{\rm o}C$ case). The relaxation in SiMC continues
even at high temperatures, but its extent is clearly smaller.

\begin{figure}  
\centering
\resizebox{0.47\textwidth}{!}{\includegraphics{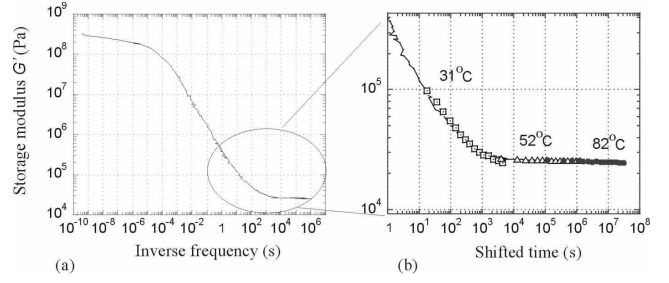}}
\caption[]{Stress relaxation analysis for the SiH material. Graph
(a) shows the DMA Master Curve, from
Fig.~\ref{time-temperature1}(a), plotted against the inverse
frequency $\omega^{-1}$. The region of long times is expanded in
Graph (b), where the real relaxation data from
 Fig.~\ref{stress relaxation curve}(a)
are shifted along the time-axis to superpose on the same Master
Curve. The shift factors required for this superposition are shown
as square symbols in Fig.~\ref{time-temperature1}(b). }
\label{masterH}
\end{figure}

\begin{figure}  
\centering
\resizebox{0.47\textwidth}{!}{\includegraphics{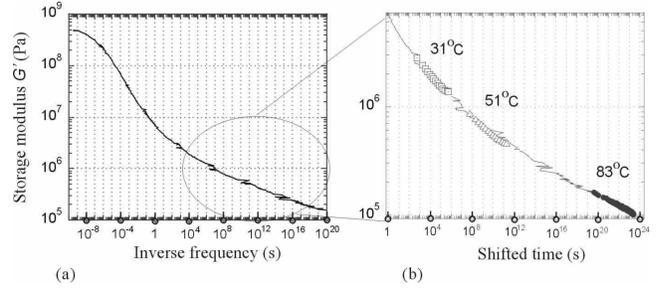}}
\caption[]{Stress relaxation analysis for the SiMC material. Graph
(a) shows the DMA Master Curve, from
Fig.~\ref{time-temperature2}(a), plotted against the inverse
frequency $\omega^{-1}$. The region of long times is expanded in
Graph (b), where the real relaxation data from
 Fig.~\ref{stress relaxation curve}(b) are shifted to superpose
on the same Master Curve. The shift factors required for this
superposition are shown as square symbols in
Fig.~\ref{time-temperature2}(b). } \label{masterMC}
\end{figure}

Instead of presenting and discussing the individual relaxation
fits and their relations, we shall follow a different route of
analysis. If we assume that the Master Curves, shown in
Figs.~\ref{time-temperature1}(a) and \ref{time-temperature2}(a),
are genuine and represent the complete relaxation dynamics in our
complex polymeric systems, then we should expect that
\underline{the same} laws must govern the static stress
relaxation. Let us plot such a Master Curve against the inverse
frequency, which would be a measure of real time. Such plots,
Figs.~\ref{masterH}(a) and \ref{masterMC}(a) look reasonable, as
one would expect the relaxing modulus to behave in real time, in
double-logarithmic scale. Of course, one cannot assign anything
more than a qualitative meaning to this representation because,
let us not forget, the $G^*(\omega)$ and $G(t-t')$ are related by
the Fourier transformation and merely mapping $G(\omega)$ onto the
inverse frequency is not a rigorous operation. It does, however
work for the power-law dependence: from the basic definitions
\begin{eqnarray}
\sigma(t) &=& \int_{-\infty}^t G(t-t') \frac{d}{dt'} \vep(t') \,
dt'  \ ;  \label{linear-def} \\
\sigma(\omega) &=&   {\rm i}\omega  \, G^*(\omega) \, \vep(\omega)
\ ,  \nonumber
\end{eqnarray}
and assuming the linear response function to decay as a power law,
at long times $G(t) \approx G_0 t^{-a}$, one can easily calculate
the corresponding estimate of the complex shear modulus:
 \begin{eqnarray}
G^*(\omega) \approx G_0 \Gamma (1-a) \left( e^{\frac{1}{2}{\rm i}
a \pi} - e^{-\frac{3}{2}{\rm i} a \pi} \right) \,  \omega^a .
\label{Gpower}
 \end{eqnarray}
Apart from a constant coefficient, this is directly the inverse
function of $G(t)$.

Let us now take the real-time stress relaxation data sets and
attempt the time-temperature superposition on them, building a
Master Curve by shifting along the time-axis. This operation works
for all temperatures, but we only present three data sets in each
of Figs.~\ref{masterH}(b) and \ref{masterMC}(b), to allow a clear
view of the underlying line, which is the DMA Master Curve for
inverse frequencies. The matching is quite remarkable, especially
if we take a look at the shift factors (now for the time-shifts,
and so with the opposite sign!) required to bring the relaxation
data onto this Master Curve. Figures~\ref{time-temperature1}(b)
and \ref{time-temperature2}(b) show the frequency shift factors
and their WLF analysis, but also the square symbols that represent
the (inverse) time shift factors for the static relaxation data.
(Recall that the Master Curves were built for the reference
temperature of around 20${}^{\rm o}C$). The agreement and the
trend in these respective shift factors is not bad.

\section{Discussion}

First of all, we must recognise the difference of our results from
some literature data. Gallani et al. and Weilepp et al.
\cite{gallani2,weilepp2} were both pointing out that the
nematic-isotropic transition is trivial for the dynamic mechanical
measurements. The reason could be that polydomain nematic
elastomers were used, although we have earlier seen the effect of
dynamic soft elasticity in a variety of polydomain elastomers
\cite{clarke01}. The drop in $G'$ below $T_{ni}$ is smaller in
polydomain samples, but it only disappears in the monodomain
$\bm{V}$-geometry, as in fact the more recent study \cite{stein01}
confirms. It is also interesting that many authors, e.g.
\cite{gallani2,fourmaux-demange}, reported their success in
time-temperature superposition across the nematic-isotropic phase
transition, which of course is related to the absence of
pronounced soft drops in $G'$ values. We can only suppose that a
different synthetic procedure of their materials led to the very
non-soft (highly semi-soft) nematic networks.

Returning to our own results, the two main results of this work
are the building of Master Curves and identifying the temperature
regimes where the superposition follows the WLF and where it fails
altogether, and establishing the direct analogy between the time
and the frequency domains. In the first part, the most important
observation is the substantial difference in rubber-plateau
(low-frequency) moduli between the nematic and isotropic phase, in
the shear geometry when the director rotation is induced. This
effect of dynamic soft elasticity can lead to a number of
technological advances, e.g., in the area of mechanical vibration
damping \cite{clarke01} (exploiting the record-high values of the
loss factor over a broad range of temperatures and frequencies),
or even suggesting a novel field of polarised acoustics (using the
fact that only certain, soft, shear-wave geometries are strongly
attenuated).

Comparing the results of static stress relaxation with those of
dynamic mechanical measurements is intriguing. Regarding the
ability to quantitatively predict the static relaxation behaviour
from the independently (and relatively quickly) acquired DMA data,
or vice versa, by inverting the Master Curve at a specific
reference temperature, we have to remain cautious. Although the
possibility is highly attractive and the procedure has worked
spectacularly well in our case, one needs to verify it on
different materials, including a check of how the classical
isotropic rubber behaves in this respect.

In our case, the Master Curve inversion has been successful and
predicts that, for instance, a ``traditional''
\cite{kupfer,kupfer2} siloxane side-chain nematic elastomer SiH,
crosslinked in a way that allows the soft elasticity, would be
able to relax under a constant strain and reach its mechanical
equilibrium (at least approximately) after some $10^4$s at room
temperature, with a very little residual relaxation after that.
This is somewhat different from the results \cite{clarkePRL} on a
very similar, but polydomain nematic elastomer, where the
relaxation have not reached its end even after extrapolation to
astronomical times. Analogous continuing relaxation was reported
in other, chemically very different polydomain nematic rubbers
\cite{hotta_terentjev2}. The effect of internal mechanical
constraints imposed by the domain interfaces is clearly dominating
long-time regime, but is absent in our present monodomain network.
In contrast, the composite nematic elastomer SiMC, albeit
monodomain, also shows a non-ending relaxation. We assume, by
analogy with the previous work on relaxation in non-crosslinked MC
polymer \cite{elias}, that internal constraints on ``free'' stress
relaxation are now imposed by the hairpin folds of long nematic
main-chain polymer strands.

In summary -- dynamic-mechanical analysis have been performed
using two different monodomain nematic elastomers having different
degrees of anisotropy and internal microstructure. Master Curves
have been built between the glassy state and the nematic-isotropic
phase transition, where the modulus saturates on the low-level
soft plateau. Above $T_{ni}$ the modulus rises substantially,
since the internal relaxation is no longer able to reduce the
elastic response -- and the further time-temperature superposition
fails. The dynamics of these elastomers is dominated by power laws
(non-universal, depending on the material and temperature), which
is confirmed by the successful operation of the Master Curve
inversion to describe the static stress relaxation. Power laws (in
time) of stress relaxation match very well with the corresponding
frequency dependence of the dynamic modulus in the appropriate
range of temperatures.

\subsection*{Acknowledgements}
We thanks EPSRC and Bridgestone Corporation for funding this work,
and Wacker Chemie for the donation of the platinum catalyst. We
are indebted to H. Finkelmann for the advice and assistance in
chemical synthesis and to A.R. Tajbakhsh for the preparation of
monodomain LCE materials.


\end{document}